# Damage in porous media due to salt crystallization

*Noushine Shahidzadeh-Bonn*[1], *Julie Desarnaud*[1,2], *François Bertrand*[1], *Xavier Chateau*[1], *Daniel Bonn*[2,3],*

1-Laboratoire Navier (umr 8205), Université Paris-Est, 2 allée Kepler, 77420 Champs-sur-Marne, France

2-Van der Waals-Zeeman Instituut (WZI), Universiteit van Amsterdam, Valckenierstraat 65, 1018 XE Amsterdam, The Netherlands.

3-Laboratoire de Physique Statistique de l'ENS (umr 8550), 24 rue Lhomond 75231 Paris cedex 05, France.

E-mail address :bonn, chateau, bertrand, desarnaud@lcpc.fr, bonn@lps.ens.fr

CORRESPONDING AUTHOR E-mail : bonn@lcpc.fr

ABSTRACT. We investigate the origins of salt damage in sandstones for the two most common salts: sodium chloride and sulfate. The results show that the observed difference in damage between the two salts is directly related to the kinetics of crystallization and the interfacial properties of the salt solutions and crystals with respect to the stone. We show that for sodium sulfate, the existence of hydrated and anhydrous crystals and specifically their dissolution and crystallization kinetics are responsible for the damage. Using MRI and optical microscopy we show that when water imbibes sodium sulfate contaminated sandstones, followed by drying at room temperature, large damage occurs in regions where pores are fully filled with salts. After partial dissolution, anhydrous sodium sulfate salt present in these regions gives rise to a very rapid growth of the hydrated phase of sulfate in the form of clusters, that form on or close to the remaining anhydrous micro crystals. The rapid growth of these clusters generates stresses in excess of the tensile strength of the stone leading to the damage. Sodium chloride only forms anhydrous crystals that consequently do not cause damage in the experiments.

INTRODUCTION.

Salt weathering is a major cause of deterioration of rocks and building materials. Many ancient structures such as the Valley of the Kings (Egypt) [1] and the Petra monument (Jordania) have been partially destroyed by salt attack [2]; also many modern constructions suffer damage [3]. Salt weathering affects buildings, engineering structures, rock outcrops and minerals within the soil profile and there is compelling evidence that its influence will increase due to the global climate change and human impacts [2,3] Since building materials are porous, any increase in soil moisture will result in greater salt mobilization and crystallization during drying, leading to damage. The salts can be naturally present in the stones, get trapped inside the porous material for instance by imbibition with salt-containing precipitation (acid rain) or ground water; it can in addition be present in the mortars used for construction.

The action of salts on natural and building materials has attracted a lot of attention over the past few decades, but remains incompletely understood [3-16]. Notably, the precise mechanisms of why certain salts (e.g., sodium sulfate) cause more damage than others (e.g., sodium chloride) under the same environmental conditions are still under debate. Several authors have attempted to explain the damage caused by sodium sulfate, based on laboratory tests of constant capillary rise (the stone is in contact at its base with the saline solution and a continuous capillary rise compensates the water evaporation) [4,7] or accelerated durability tests (wetting /drying cycles with saturated salt solutions followed by drying in an oven at around 100°C) [6]. In these papers, different parameters were studied: the impact of the ion concentration (notably the formation of supersaturated solutions [2,5,9]), the structure of the precipitated salt [14,15] and the effect of the pore size distribution of the stone [7,8]. However, the relevance of these accelerated tests to judge the durability of building stone has been questioned [6,14].

For the specific case of sodium sulfate, two crystalline phases exist and it is not yet clear which phase, hydrated (stable decahydrate, metastable hepta hydrate) or anhydrous, is the one responsible for the damage. Rodriguez Navarro *et al.* [4] argued that the precipitation of thernardite (anhydrous salt) could be responsible for the damage observed after 30 days of constant capillary rise test at low relative humidities (35%). This test imposes a large evaporation rate and there is possible precipitation of both anhydrous and hydrated phases (observed by ESEM), which should be compared to experiments done at high relative humidity (60%) for which only the hydrated phase is observed. In their experiments no reduction in porosity of the limestones in the presence of sulfate either at high (60%) or at low (35%) relative humidities for sodium sulfate was observed. Tests with sodium chloride, on the other hand, showed a reduction of the porosity by half, although less damage occurs. In another series of experiments, based on cyclic impregnation with saline solutions and fast drying (at 100°C), Tsui et al [6] by changing the temperature of the saline solution for impregnation (under and above mirabilite stability which 32°C) have shown that damage occurs at 20°C after several cycles. They concluded that when enough salt is present damage occurs due to the precipitation of mirabilite at the wetting step after the dissolution of thernardite which gives highly superstaturated solutions.

From the theoretical side, different explanations for salt weathering have been proposed, the most popular being the crystallization pressure of supersaturated salt solutions [5-11]. Although the supersaturation can account for an excess pressure exerted by the salt crystals against the pore walls [8-11], this does not explain the difference between different salts;

notably for the two archetypical examples sodium chloride and sulfate, the former has a higher crystallization pressure but the latter causes much larger damage [5,9].

Consequently, a detailed understanding of how crystal growth within the porous media leads to the damage still remains elusive. What is clear is that some salts preferably crystallize on the exterior wall of the stones, a phenomenon called efflorescence, whereas others prefer to crystallize within the porous medium, called subflorescence. It is well known that the latter is far more damaging than the former. However, specifically for sodium chloride and sodium sulfate the situation is not very clear. Some authors have reported more efflorescence in the case of chloride whereas others mention more efflorescence in the case of sulfate contaminated stones [4,7,10,14]. Evaporation of solutions of sodium chloride and sodium sulfate as droplets and in square microcapillaries have shown that the combination of interfacial properties of the different crystalline phases with transport properties of the liquid both governs the crystallization dynamics and determines whether the salt will effloresce or subfloresce during the drying[16].

Here, we investigate the behaviour of salt contaminated stones, for sodium chloride and sodium sulfate subject to realistic exposure conditions. In order to simulate realistic exposure conditions and to see how damage will occur at the macroscopic scale we follow the behaviour of salt contaminated sandstones (obtained by one cycle of impregnation and drying with saturated salt solution) once rewetted with pure water and dried under constant environmental conditions (T=21°C and RH~45%). At the microscopic scale we have followed (re-)wetting and drying using optical microscopy. In these experiments, we have followed the dissolution and have measured the speed of crystallization growth during rewetting and drying of dried salt solution droplets on flat hydrophilic surfaces and in salt contaminated square capillary tubes as simple model systems for a single pore within a porous medium contaminated with salt. We find that the observed difference in damage between the two salts is directly related to the kinetics of crystallization and the existence of hydrated and anhydrous crystals. Using MRI we show that when water imbibes sodium sulfate contaminated sandstones, large damage occurs in regions where pores are fully filled with salts. After partial dissolution, anhydrous sodium sulfate salt present in these regions gives rise to a very rapid growth of the hydrated phase of sulfate in the form of clusters, that form on or close to the remaining anhydrous micro crystals. The rapid growth of the clusters generates stresses in excess of the tensile strength of the stone and hence lead to the damage. Sodium chloride only forms anhydrous crystals that consequently do not cause damage in the experiments. To our knowledge, the direct evidence of the difference in crystallization dynamics between drying and rewetting has not been reported in the literature so far.

EXPERIMENTAL

We investigate crystallization of $Na_2SO_4$ and NaCl during wetting and drying cycles in a porous sandstone (of porosity $\phi \sim 25\%$, and pore diameter $d_p \sim 30$ μm) both at macroscopic (MRI measurements, weight balance) and microscopic scales (crystallization growth under the microscope). Sodium sulfate has two stable crystal phases at room temperature: the anhydrous phase ($Na_2SO_4$, thenardite, saturation concentration 35 wt%, or $C_{sat}=3.8$ mol/L (M) at 21°C) and the decahydrated form ($Na_2SO_4 \cdot 10H_2O$, mirabilite, 16.5 wt%, $C_{sat}=1.4$M, 21°C) [5,16]. In addition, a metastable heptahydrated phase has been reported in some laboratory experiments [11,13,18]. Saturated solutions of sodium sulfate were prepared by dissolving thenardite (anhydrous crystals) at room temperature till mirabilite precipitate in the bottle (16.5 wt%, $C_{sat}=1.4$M, 21°C).

The relative stability of mirabilite and thenardite in open-air depends on temperature and relative humidity. X-ray crystallography experiments on our samples show that even if mirabilite is formed initially, if left in contact with air at relative humidity RH~ 40% this will transform into the anhydrous phase (thenardite). So in contact with (dry enough) air, thenardite is the thermodynamically stable state. Sodium chloride has only a stable anhydrous phase (halite, 26.4 wt%, $C_{sat}$=6.1mol/L, 21°C). The surface tensions of the different solutions ($\gamma_{water}$~71.7mN/m, $\gamma_{Na2SO4}$~77mN/m, $\gamma_{NaCl}$~84 mN/m) were measured using the drop weight technique. All experiments were done at 21°C and RH~45%±4.

For the microscale experiments, rectangular glass capillaries (100x800 µm) were used as a simple model system for a single pore in a porous medium. To avoid gravity effects, experiments (wetting and drying cycles) were done in horizontal capillaries under a microscope.

Droplets (20 µl) of saturated salt solutions (NaCl and $Na_2SO_4$) were evaporated on hydrophilic glass slides (Corning) (Cycle 1, C1) and observed under an optical microscope. The salt crystals (Cycle 2, C2) obtained after drying were rewetted with pure water (same volume of water as in C1). The direct imaging experiment allows us to assess the kinetics of crystal dissolution and growth during cycle 1 and 2 .

For the macroscale experiments, samples of Prague sandstones ($\phi$~25%, $d_p$=30 µm) were saturated by imbition at ambient temperature (21°). For the first wetting/drying cycle, imbition was done with the saturated salt solutions and for the second cycle only with pure water. The samples were dried under controlled laboratory conditions (T=21°C, RH=45±4%). The weight of stones is followed in time during drying on an automated balance with a precision of (±0.001 g).

Proton MRI was used to follow the dynamics of desaturation within the stone during drying cycles; experiments were performed on a vertical Bruker Spectrometer [18]. Water (proton) density profiles were obtained along the z-axis (height). The field strength is 0.5T for the experiments described here, and the MRI sequence is a 1d spin echo, to minimize the echo time and possible differences in relaxation time due to concentration gradients. The measured

NMR relaxation times were for pure water T1=2.541s and T2=2.046s, for a saturated Na2SO4 solution T1=1.94 s and T2=1.64s, and for saturated Na2SO4 solution in equilibrium with hydrated crystals: T1=1.94s and T2=98.42 ms. The typical reproducibility of such measurements was found to be better than 5%. The weight of the sample is measured prior to each MRI measurement.

RESULTS

During a first cycle (C1), sandstones samples were saturated by imbibition with saturated salt solutions (16.5 wt% for $Na_2SO_4$ or 26.4 wt% for NaCl) and left to dry.

At the end of drying, the amount of salt crystallized at the surface as efflorescence is gently removed with a toothbrush. The efflorescence is ~ 50%±5 of the initial mass of salt in the case of $Na_2SO_4$ and ~ 10%±2 for NaCl. This result shows that the saturated sodium sulfate solution clearly gives more efflorescence during drying at room temperature than sodium chloride under the same environmental conditions. For the sodium chloride deposit , as was also found by Lubelli et al [14] , we observe a stronger adhesion to the sandstone. To the contrary, for sodium sulfate, the efflorescence deposit can be removed very easily. No damage (loss of material) is observed, neither for the sulfate nor for the chloride once the efflorescence part is removed.

In a second cycle (C2), the salt containing stones (for sodium sulfate 50%±5 and for sodium chloride 90%±5 of the initial mass of the salt remains in the stone) are saturated with pure water and left to dry under the same environmental conditions as in C1. At the end of drying of C2 for sodium sulfate again a large fraction of the salt crystallizes at the surface, as efflorescence, where as in the case of sodium chloride the amount of salt as efflorescence remain small (Fig.1). During the gentle removal of the efflorescence, a significant disintegration of the stone can be observed in the case of sodium sulfate. The percentage of damage is assessed by gently removing the efflorescent salt, washing out the remaining salt, drying the stone and weighing the stone; then $Damage = \left( M_{initial}^{dry\ stone} - M_{final}^{washed\ stone} \right) \bigg/ M_{initial}^{dry\ stone} \times 100\ \%$. The damage can be up to 10-12% of the initial mass of the stone. For sodium chloride, although a larger amount of salt was present in the stone, no measurable damage occurred at this stage (Fig 1b and c).

These experimental observations imply that the cause of damage can not simply be due to the crystallization during evaporation of saturated salt solutions in the porous medium, which also happens during C1.

Magnetic Resonance Imaging (MRI) was used to follow the dynamics of drying within the stones for each cycle [19]. The MRI measurement of the amount of water contained in the sulfate-containing stones reveals that there is about 5% less water than can fit in the porous space at the end of the imbibition for C2 (Fig. 2). This amount corresponds exactly to the volume of sulfate present as subflorescence in the porous medium and, in addition, it is precisely located where later the damage is observed, i.e. in the upper part of the sample (Fig. 1b). This follows from the amount of sodium sulfate salt left in the porous stone after C1 ($M_{salt}^{subflorescence}$ =0.46g, corresponding to $V_{salt}$=0.17cm$^3$ of thenardite $\rho$=2,66 g/cm$^3$), which is 4.6% of the pore volume of the sample (15cm$^3$ with a porosity of $\phi$=25%).

Another key result from the MRI data is that in the presence of sodium sulfate, the water loss during the second cycle of drying appears to happen faster than in C1 (see profiles after 5h in Fig. 2), although experiments were done under the exact same environmental conditions. No receding front is observed on the profiles till the end of drying. In addition, there is a serious discrepancy between the rate of evaporation obtained from the MRI profiles and weight measurements done simultaneously on the same sample (Fig. 2(b)). Such a discrepancy is not observed for sodium chloride. At first sight, these observations are puzzling, but they can be explained by noting that the 'missing' water does not disappear, but rather goes into hydrated crystals that are not visible with the MRI. Indeed, an independent MRI measurement of hydrated crystals of sodium sulfate in a beaker shows that the water in those crystals does not contribute to the measured proton density, because their NMR relaxation time becomes very short; they do of course contribute to the measured weight.

Thus, the hydrated phase of sodium sulfate starts to grow in the porous network and the difference between the MRI and weight measurements allows to calculate the amount of hydrated crystals.

Here we consider only the stable decahydrated phase [11,13] although a metastable heptahydrate phase may form ; however, this phase has been reported not to provoke damage [11] and thus can not be responsible for the disintegration of the stone. Experiments on whether a heptahydrate may form during the processes described here are in progress, but are beyond the scope of this paper.

From the measurements, we find that after 5 hours during C1, hydrated crystals constitute about 64% of the sulfate present; however no damage is observed. Afterwards, these hydrated crystals lose their water (e.g., the amount of hydrated phase decreases to 42% after 22 hours; and C2 is started only after all of the water has disappeared) and transform into anhydrous crystals at the end of drying.

The MRI data show that in Cycle 2, the anhydrous sulfate is again transformed into its hydrated form (about 80% of the sulfate present after 5 hours is hydrated) which thereafter leads to severe damage. Although the formation of the hydrated crystals (mirabilite) can be considered as responsible for the damage (as proposed already in [5,6]), it can not explain why damage only occurs in C2 in these experiments; i.e. when pure water is added to salt contaminated stone. It should be noted that the global quantity of salt present as subfloresence at this stage is roughly half that initially present. Consequently, the key must lie in understanding how the anhydrous crystals are transformed into the hydrated phase when put in contact with water in C2.

To investigate this, the dynamics of crystallization for the two salts were followed at the microscopic scale. The crystallization was studied in rectangular (100x800μm) glass capillaries, as a simple model for a pore in the sandstone. Following the same procedure as for the macroscale experiments, capillaries were saturated with salt solutions (C1) and after drying, the salt-containing capillaries were rewetted with pure water (C2) followed by a second drying step.

During C1 and for sodium sulfate solutions, we observe the direct formation of hydrated crystals in the capillary (Fig.3a) as is evident from their crystal shape [15, 16]. At the end of drying, the hydrated crystals transform into anhydrous ones, and thus lose their water, as confirmed by weight measurements. However the macroscopically observed form of the crystals remains the same so that it is likely that they transform into small anhydrous micro crystals. For C2, the latter are rewetted with the same volume of water that was initially present. Upon rewetting, we observe an only partial dissolution of the anhydrous crystals. Due to the lower solubility of the hydrated phase compared to the anhydrous one, a region of highly concentrated solution (with respect to hydrated phase) is developed once part of the crystallites start to dissolve. Subsequently, the small thenardite crystallites that remain present in the solution are observed to act as seeds (nucleation sites) for the rapid growth of large amounts of hydrated crystals giving rise to structures that bear a similarity to a bunch of grapes (Fig. 3); these clusters are observed to expand much more rapidly than the growth of hydrated crystals during C1. As far as the resolution of the microscope allows us to see, the newly formed hydrated crystals indeed form onto the existing thenardite seeds. It may however also be that they form in the immediate vicinity of the thenardite microcrystals, since the supersaturation is higher there.

The growth velocity of crystals was determined from the microscopy images during each cycle since it is directly related to the supersaturation $C/C_{sat}$. For C1, the growth speed is about 0.022 µm/s, in the fastest growing crystalline direction (Fig.4). However, during C2 the growth velocity of the hydrated crystals in the grape structure is found to be more than an order of magnitude larger: 0.26 µm/s ( Fig.4, inset).

It was shown experimentally [21] that the growth rate of crystals increases linearly with increasing supersaturation for sodium sulfate. The slope was found to be 1.2 µm s$^{-1}$/(mol l$^{-1}$), which allows to calculate the supersaturation in our experiments from the growth rate. We find a supersaturation of $\approx 0.018$ mol/L in C1, and $\approx 0.21$ mol/L in C2. Are then these supersaturations sufficient to generate stresses that exceed the tensile strength of our material? We have determined the tensile strength of the sandstone using a three point flexion experiment [22], measuring the maximum force the sample can withstand without breaking. We find ~0.9 MPa, in good agreement with previous measurements done on the same material [23].

The crystallization pressure as a function of the supersaturation can be calculated as $\Delta P = RT/V_m \ln a/a_0$, with $V_m$ the molar volume and $a$ and $a_0$ the solute activities of the saturated and supersaturated solution [9,12]. A crystallization pressure of about $\Delta P \sim 6.5$ MPa per mole l$^{-1}$ of supersaturation of hydrated crystals (mirabilite) is estimated [9] for small supersaturations which pertains to our experimental conditions. Thus during C1, in a supersaturated solution at 0.018 mol/L, the crystallization pressure of mirabilite during growth is $\Delta P = 0.12$ MPa, much smaller that the tensile strength of the sandstone (~ 0.9 MPa), explaining why there is no damage at this stage. On the other hand, in C2, the supersaturation of 0.21 mole l$^{-1}$ gives rise to a crystallization pressure of $\Delta P = 1.4$ MPa, this time sufficient to damage the sandstone, as is indeed observed experimentally.

For sodium chloride the microscale experiments reveal that the kinetics of dissolution/crystallization growth of anhydrous cubic crystals are the same, to within the experimental accuracy, in both C1 and C2. Moreover, the speed of growth of sodium chloride crystals (0.027 μm s$^{-1}$) is found to be of the same order as that of the slow growth of hydrated sodium sulfate during C1. This measured crystallization velocity leads to an inferred supersaturation of 0.018M [25]. Using the formula above, this leads to a crystallization pressure of ΔP =0.3 MPa, again not sufficient to break the stone. This therefore solves the issue why sodium sulfate causes damage, whereas sodium chloride does not in our experiments.

The drop evaporation (see [24].for a detailed account of the evaporation of aqueous drops) and rewetting experiments shed some light on the difference between subflorescence and efflorescence for the two salts (Fig.5). As was explained previously[16], when droplets of saturated salt solutions (NaCl and Na$_2$SO$_4$) are deposited on a hydrophilic glass slide, the crystallization starts at the liquid/air interface enhancing the spreading power of the solution by lowering the interfacial energy γ$_{lv}$ during crystallization and evaporation (Fig. 5a). Now, if the salt crystals deposits are rewetted with pure water (same volume than C1) an only partial dissolution of the salt can be observed under the microscope, as the droplet evaporates at the same time. For NaCl cubic crystals start to grow again, both as new crystals and on old partially dissolved crystals, giving a crystals deposit roughly at the same location than in C1. On the other hand, for sodium sulfate after a partial dissolution and as soon as crystallization starts again at the interface, the solution spreads out tremendously due to the decrease of liquid surface tension leading to the crystallization in the spreading film. The Sodium sulfate solution is less concentrated and has lower liquid/vapor interfacial tension (77 mN/m) than the sodium chloride solution (84 mN/m). Consequently when crystallization starts, the surface tension decreases more rapidly to the value of 71 mN/m and provoking the complete spreading of the solution. The consequence is that sodium sulfate is transported due to the wetting film from one location to the other during dissolution and crystallization, as it can be observed in Figure 2C. Based on these results and by taking into account the hydrophilic character of the sandstone, a similar phenomenon should occur in our macroscale experiments on sandstones explaining on the one hand why more sulfate is able to move out of the stone, and cause efflorescence and on the other hand why no drying front is observed on MRI profiles until the end of drying [19].

DISCUSSION AND CONCLUSION

The large damage that crystallization of sodium sulfate causes has been known for centuries, and has even been used in the past to evaluate the strength of building materials . It was also believed that the large difference between sodium sulfate and a less damaging salt such as sodium chloride was due to the fact that the sulfate can form hydrated crystals [5,6,9,12,20]. The detailed experiments [4, 5, 9.] even provided evidence that the larger damage caused by the sulfates was due to a larger supersaturation in the sulfate. Such a larger supersaturation leads to a higher crystallization pressure through the formula given above, and hence can account for the larger damage. However what is less clear from their experiments is how this larger supersaturation is obtained, and hence, under which conditions it will occur, a matter of paramount importance if one wants to prevent the damage. What we show here is that salts such as sodium sulfate with hydrated and anhydrous phases can lead to severe damage in porous materials during rewetting /drying cycles because of the only partial dissolution of anhydrous crystals in regions (pores) that are highly concentrated in salt. The thenardite microcrystals dissolve very rapidly, and in part act as seeds to form large amount of hydrated crystals creating grape-like structures that expand rapidly. This explains the higher supersaturation of sulfate solutions in comparison with chloride. Subsequently, the clusters of crystals generate stresses larger than the tensile strength of the stone, which then leads to the damage. All these effects are related to the existence of both hydrated and anhydrous forms and the large difference of saturation concentration of the salt with respect to the hydrated and anhydrous crystals. Consequently, they are absent in salts such as sodium chloride. The supersaturation was inferred directly the in situ measurement of the growth rate of the crystals, allowing for the measurement of the actual supersaturation in the vicinity of the growing crystals.

Comparing to the measurements of [5,6,11], the new insight from our MRI experiments is that although hydrated crystals form during the first cycle, no damage is observed. We thus find that there is no direct relation between mirabilite formation and damage, but that the way the mirabilite forms is of key importance. We also provide evidence that during wetting and drying of sodium sulfate contaminated stone, the formation of the hydrated sulfate crystals enhances the spreading power of the salt solution on the pore network of the sandstone explaining why more efflorescence is observed compare to sodium chloride..

The better understanding of the mechanisms involved in damaging developed here can significantly improve the methods used to prevent damage. For example, the use of water based poulticing methods for desalination on stones highly contaminated by sodium sulfate is counterindicated since it requires rewetting of the stones.

REFERENCES

(1) Nicolaescu, A. "Conference Review: Salt Weathering on Buildings and Stone Sculptures", *e-conservation magazine*, **2009**, 8, 6.

(2) Wust, R.A.; McLane, R.; *Engineering Geology* **2000**, 58, 163.

(3) Goudies, A.S.; Viles, HA. *Salt weathering hazard*, Wiley London **1997.**

(4) Rodriguez-Navarro, C.; Doehne, E. *Earth Surf. Process. Landforms* 1**999**, 24 191-209.

(5) Flatt, R.J. *J. Cryst. Growth.* **2002,** *242*, 435-454.

(6) Tsui, N.; Flatt, R.; Scherer, G.W. *J. Cult. Herit.* **2003**, *4*, 109-115.

(7) Scherer, G.W. *Cem. Concr. Res.* **2004**, *34*, 1613-1624.

(8) Coussy, O, *J.,* **2006***, Mech.Phys. Solids*, *54*, 1517.

(9) Steiger, M. *J. Cryst. Growth.* **2005**, *282*, 455-469

(10) Sghaier, N;Prat, M. 2009, *Transp.Porous Med.*,**2009**, *80*, 3, 1573-1634

(11) Rijners, L.A.; Huinink, H.P.; Pel, L.; Kopinga, K. *Phys. Rev. Lett.* **2005**, *94*, 75503.

(12) Steiger M., Asmussen S., *Geochimica and cosmochimica Acta*, **2008**, *72,* 4291,

(13) Espinosa Marzal, R.M. ;Scherer G.W.; *Environ Geol* **, 2008,** *56*, 605.

(14) Lubelli, B.; Van Hees, R. R. J.; Groot, C.J. W. P. Stud. Conserv. **2006**, *51*, 41-56.

(15) Rodriguez-Navarro, C.; Doehne, E.; Sebastian, E. *Cem. Concr. Res.* **2000,** *30,* 1527-1534.

(16)    Shahidzadeh-Bonn, N.; Rafai S; Bonn, D.; Wegdam, G; *Langmuir* **2008**, 24, 8599

(17)    Oswald I.D.H. et al. J.Am.Chem.Soc., **2008**, *30*, 17795

(18)    Bonn, D.; Rodts, S. ; Groenink, M.; Rafai, S.; Shahidzadeh-Bonn, N.; Coussot, P. *Annu. Rev. Fluid Mech.*, **2008,** *40*, 209-233..

(19)    Shahidzadeh-Bonn, N; Bertrand, F; Chateau, X; Bonn, D, *Poro mechancis IV* , 4th Biot Conference proceeding, **2009**, 246-251.

(20)    Chatterji, S ; Jensen, A. D. *Nordic Concr. Res.*,**1989**, *8*, 56–61.

(21)    Rosenblatt, D; Marks, S.B.; Pigford*,* R.L. *Ind. Eng. Chem. Fundamen*. **1984**,23, 143.

(22)    Shahidzadeh-Bonn, N; Vie, P; Chateau, X; Roux, JN; Bonn, D. *Phys.Rev.Lett.*, **2005**, *95*, 175501.

(23)    Pavlík, Z. ; Michálek, P. ; Pavlíková, M. ; Kopecká, I. ; Maxová, I.. *Construction and Building Materials* **2008**, *22*, 1736.

(24)    Shahidzadeh-Bonn, N.; Rafai. S.; Azouni , A.; Bonn, D. *J.Fluid Mech.* **2006**, *549*, 307-313.

(25)    Al-Jibbouri S.;Ulrich J.; *J. Cryst.Growth* **2002**, 234, 237.

FIGURES

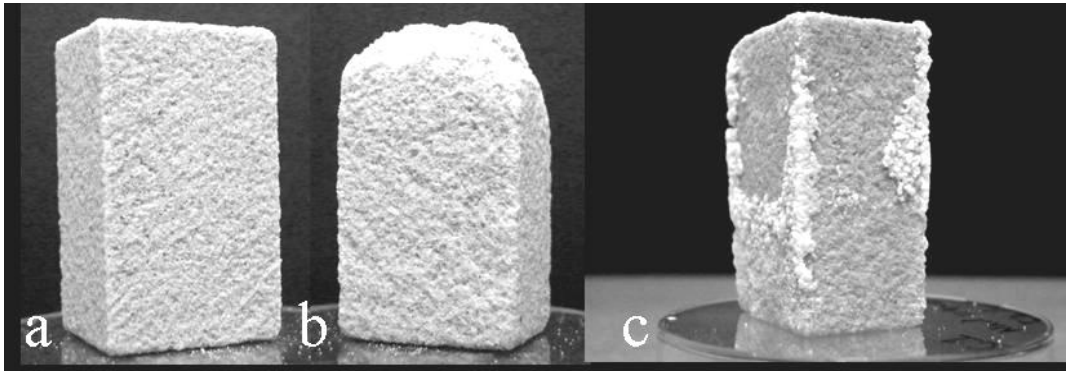

**Fig.1.** (a) Prague (Mšené) sandstone sample (2x2 x 4 cm) ($\phi \sim$ 25% and $d_p$ = 30 μm) before the experiments. (b)-(c) After two cycles of wetting/drying, imbibing with $Na_2SO_4$ and NaCl respectively in the first cycle and imbibition with pure water in the second cycle.(b) Severe damage (loss of material) in the presence of $Na_2SO_4$ (c) strong adhered efflorescence but no damage in the presence of NaCl. Experimental conditions T = 21°C, RH = 45±5%.

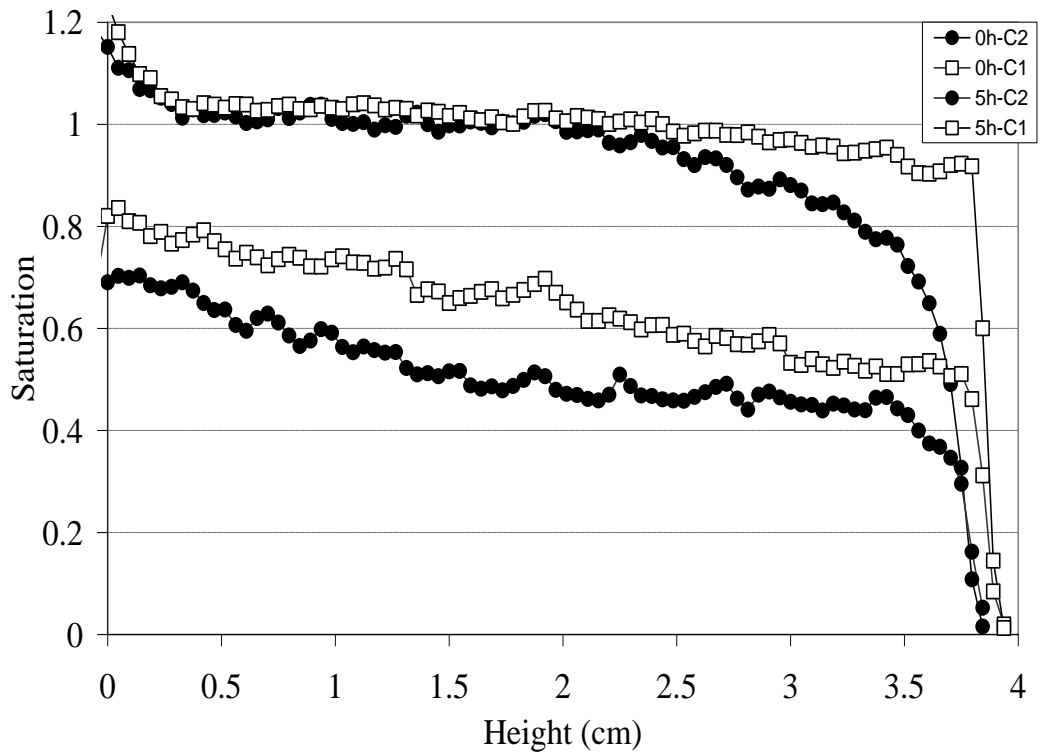

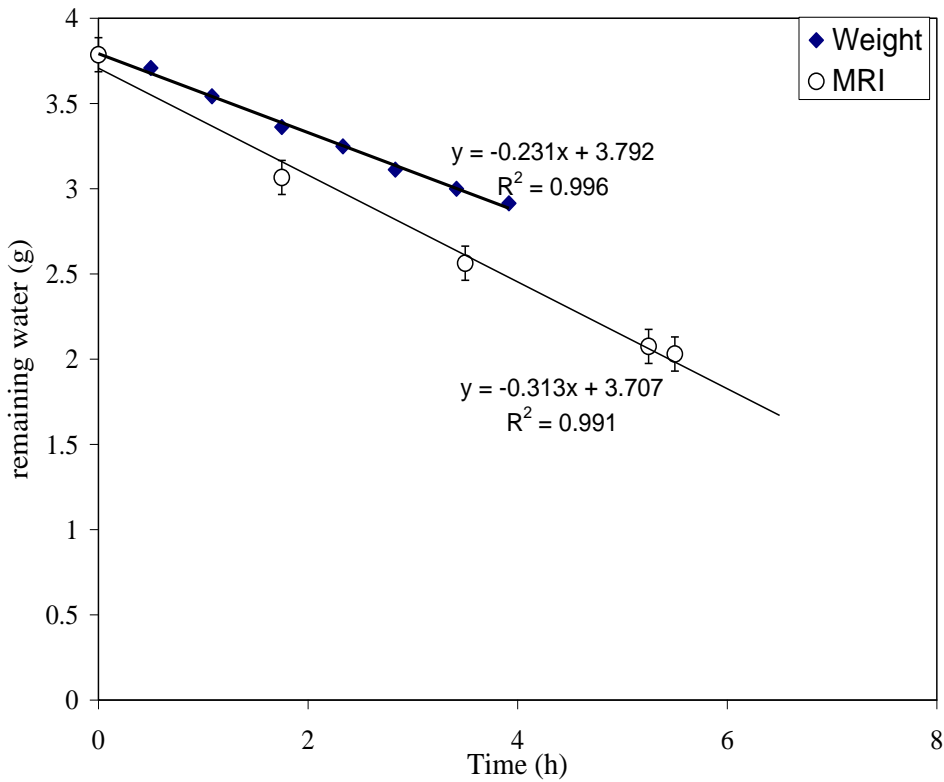

**Fig.2(a).** Comparison of MRI saturation profiles between the time instants just after imbibition and after 5 hours of drying in C1 and C2. The sample is imbibed with a saturated

sodium sulfate solution in the first cycle, and with pure water in the second one. Squares are for C1, circles for C2; **(b)** Water content of the stone obtained from the MRI measurements (open symbols) and the weight measurements (filled symbols) during cycle 2. The difference reveals the formation of hydrated crystals that are not visible as water in the MRI experiments.

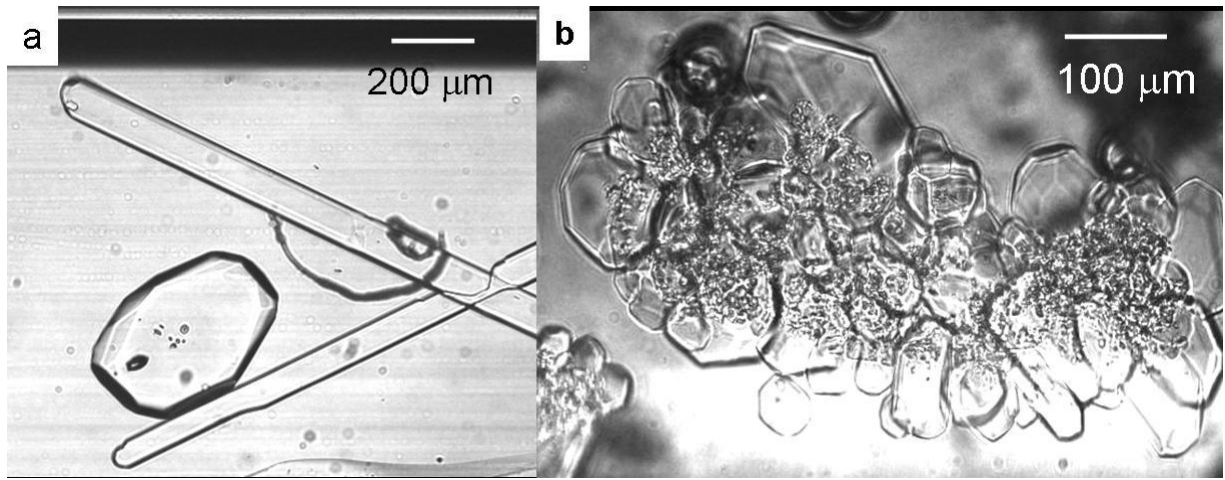

**Fig 3.** Microphotographs of the formation of hydrated crystals of sodium sulfate (mirabilite) in a rectangular micro capillaries (a) Slow growth of large crystals from solution in cycle 1 (b) After rewetting with pure water (cycle 2), the anhydrous sodium sulfate crystallites (thenardite) that are visible as the small grains in the center of the aggregate do not dissolve completely, and the hydrated crystals grow from there. The latter are visible as the facetted transparent crystals.

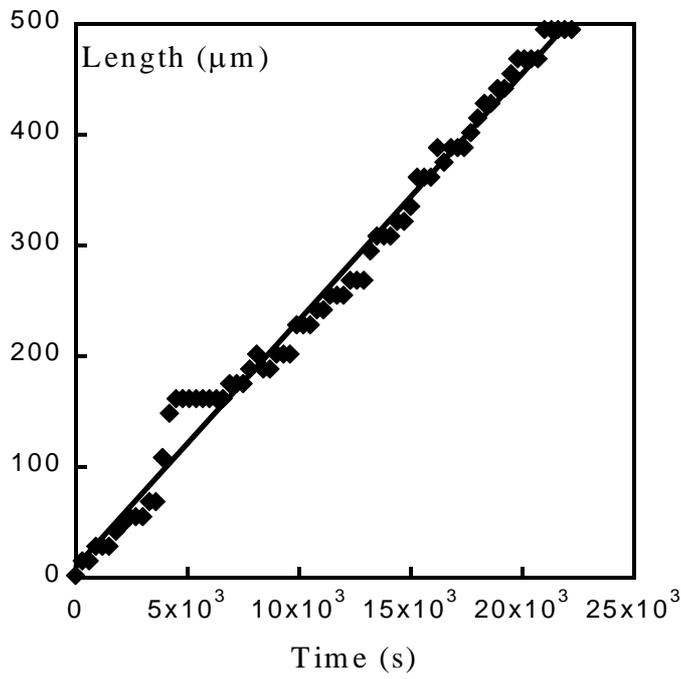

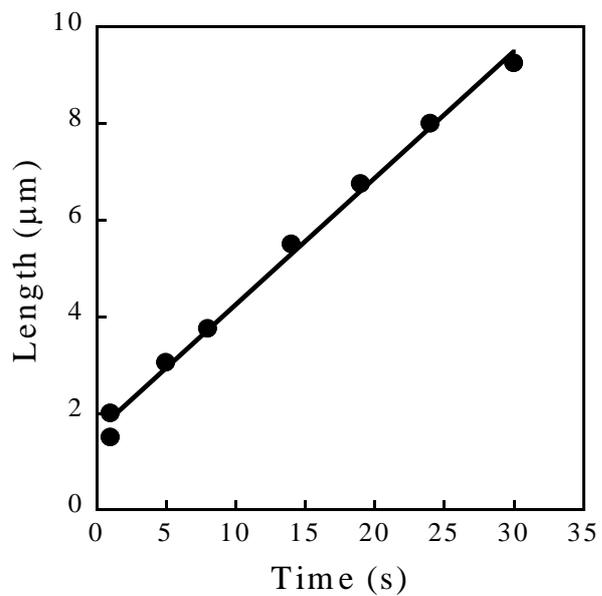

**Fig.4**. Dynamics of crystallization growth: typical size of a mirabilite crystal forming during the first cycle as a function of time in a rectangular microcapillary. Inset (to the right): the much faster growth when the mirabilite crystals form in clusters as observed in Fig.3(b).

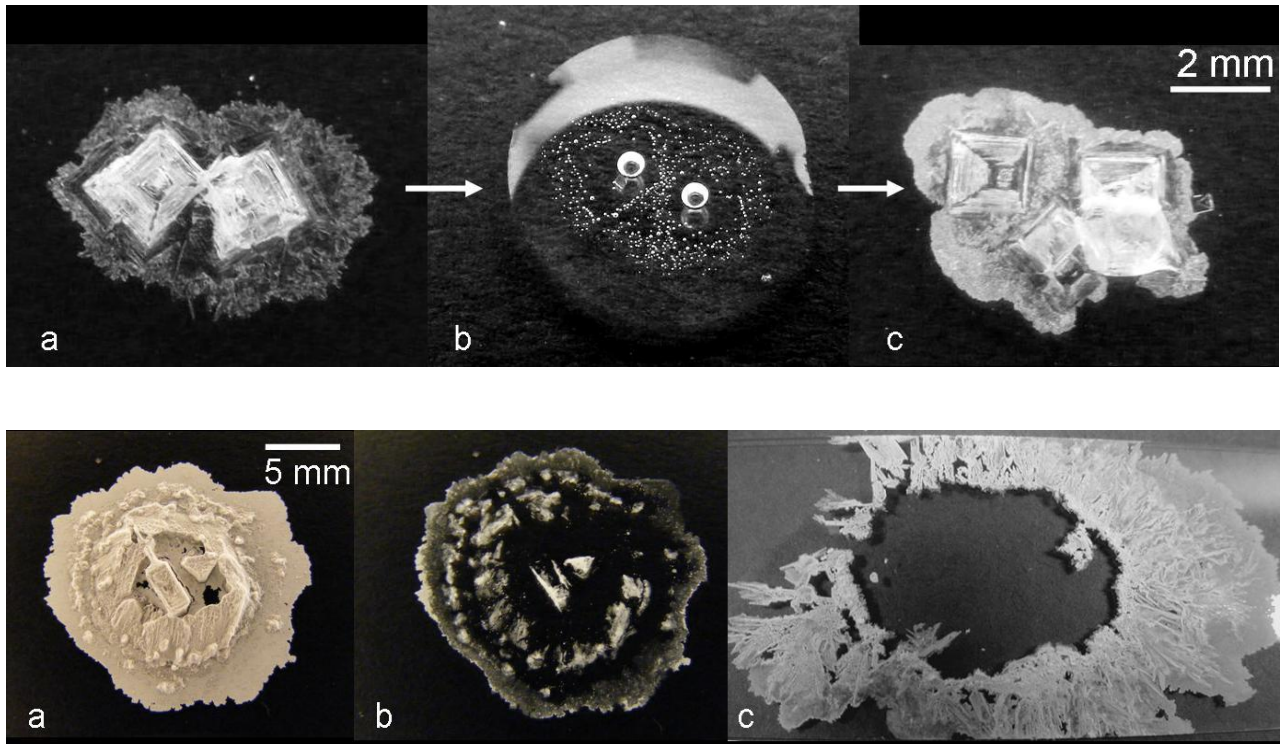

**Fig 5.** Drop evaporation and rewetting experiments, top images are for NaCl whereas bottom images are for $Na_2SO_4$. In both series, a) corresponds to the salt deposit after drop (20μl) of a saturated salt solution has dried. b) corresponds to the rewetting with the same amount of water as was initially present in the drop under a). In most cases, an only partial dissolution of the salt is observed, as the droplet evaporates at the same time. c) shows the deposit after drying a second time. For NaCl a pattern similar to that under a) is observed. However, the sulfate spreads out tremendously after rewetting, leaving a much larger area covered after drying.